\documentclass[pss]{wiley2sp} 
\usepackage{amsmath}
\usepackage{amssymb}
\usepackage{bm}             
\usepackage{verbatim}             

\begin{document}

\title{Determination of Out-of-plane Spin Polarization of Topological Surface States 
  by Spin Hall Effect Tunneling Spectroscopy}

\author{%
  Matthias G\"otte\textsuperscript{\textsf{\bfseries 1}} and
  Thomas Dahm\textsuperscript{\Ast,\textsf{\bfseries 1}}}

\mail{e-mail
  \textsf{thomas.dahm@uni-bielefeld.de}}

\institute{%
  \textsuperscript{1}\,Universit\"at Bielefeld, Fakult\"at f\"ur Physik, Postfach 100131, D-33501
Bielefeld, Germany}


\abstract{\bf%
Determining the detailed spin texture of topological surface states is
important when one wants to apply topological insulators in spintronic
devices. In principle, the in-plane spin component of the surface states
can be measured by a method analogous to the so-called Meservey-Tedrow
technique. In the present work we suggest that the out-of-plane spin component
can be determined by spin Hall effect tunneling spectroscopy.
We derive an analytical formula that allows to extract the 
out-of-plane spin component from spin Hall effect tunneling spectra.
We test our formula using realistic tight-binding models of
Bi$_2$Se$_3$ and Sb$_2$Te$_3$. We demonstrate that the extracted out-of-plane spin
polarization is in very good agreement with the actual out-of-plane spin
polarization.

 }

\maketitle   

\section{Introduction}
Topological insulators (TI) possess surface states that are protected
by the momentum space topology of the material \cite{Bernevig2006,Fu2007,Koenig2007}. An interesting
feature of these surface states is spin-momentum locking, which
means that momentum and spin of an electronic surface state
are strictly related to each other, i.e. electrons with
opposite spin propagate into opposite directions \cite{Ando2013,Bruene2012,Pan:PRL106,Wu76,MinePRL2019,HatsudaScience2018}. This feature
makes topological insulators interesting materials for spintronic
devices \cite{Tanaka,Garate,Krueckl,Goette-PRA2014,GJD,Han,Khang2018,He}. 
In particular, the combination of topological insulators
with ferromagnets promises interesting device
applications like magnetoresistance devices
\cite{Burkov,Vali,Taguchi,Goette-PRA2014} or
spin-orbit torque \cite{Fan2014,Fischer2016,Han,Khang2018}. It has been
demonstrated that topological insulators can be made ferromagnetic by
magnetic proximity effect \cite{Vobornik,Yang2014,Li2015} or doping 
with magnetic impurities \cite{Hor2010,QAHScience,Kou2014,QAHVdoped}.
Recently, the first intrinsic magnetic topological insulator has been
realized experimentally \cite{IMTI}.

In previous work we have shown that the combination of ferromagnets with 
topological insulators provides several interesting opportunities: it is
possible to construct spin current generators, detectors, and spin transistors
\cite{GJD}. One can create flat surface bands \cite{PD,PGGD} and a magnetically
induced Weyl semimetal state can be reached \cite{PGGD}.

For spintronic devices made from topological insulators it is important
to determine spin texture and degree of spin polarization of the surface 
states. Experimentally,
spin and angle resolved photoemission spectroscopy (SARPES) has been
used to measure the spin texture \cite{Hsieh2009,Souma2011,Jozwiak,Pan:PRL106}. However, the values reported
for the degree of spin polarization varied strongly \cite{Pan:PRL106,Jozwiak,Sanchez,Landolt}.
It was pointed out that the spin of the photoelectrons can be
different from the electrons in the topological surface states
depending on photon energy and photon polarization \cite{Sanchez,Park}.
Therefore, alternative techniques to determine the spin texture and degree 
of spin polarization would be valuable.

For ferromagnetic materials there exists the so-called Meservey-Tedrow
technique \cite{TedrowMeservey-PRL26,TedrowMeservey-PRB7} to measure
the spin polarization. In this
technique one uses tunneling spectroscopy from a superconducting
thin film in a strong external magnetic field.
We have recently demonstrated that the Meservey-Tedrow technique can be 
adapted to topological insulator surface states and
allows to measure the in-plane component of the surface state spin
polarization \cite{GD}. However, the method is insensitive to the
out-of-plane component of the spin texture.
 
In the present work we will show that spin Hall effect tunneling spectroscopy
can be
used to measure the out-of-plane spin polarization of the topological surface
states and thus can complement the Meservey-Tedrow technique to give a
complete measurement of the surface state spin texture by tunneling spectroscopy.
Spin Hall effect tunneling spectroscopy was introduced by Liu \textit{et al}
\cite{Liu-NatPhys2014} as a means to measure the Spin Hall effect under finite bias
voltages. This technique allows to measure the energy dependence of the
Spin Hall effect near the Fermi energy of a given metal. It was shown that
the charge-spin conversion efficiency of topological insulator surface states
can be determined experimentally using this technique \cite{LLiu2}.

\begin{figure}[t]%
\includegraphics*[width=\linewidth]{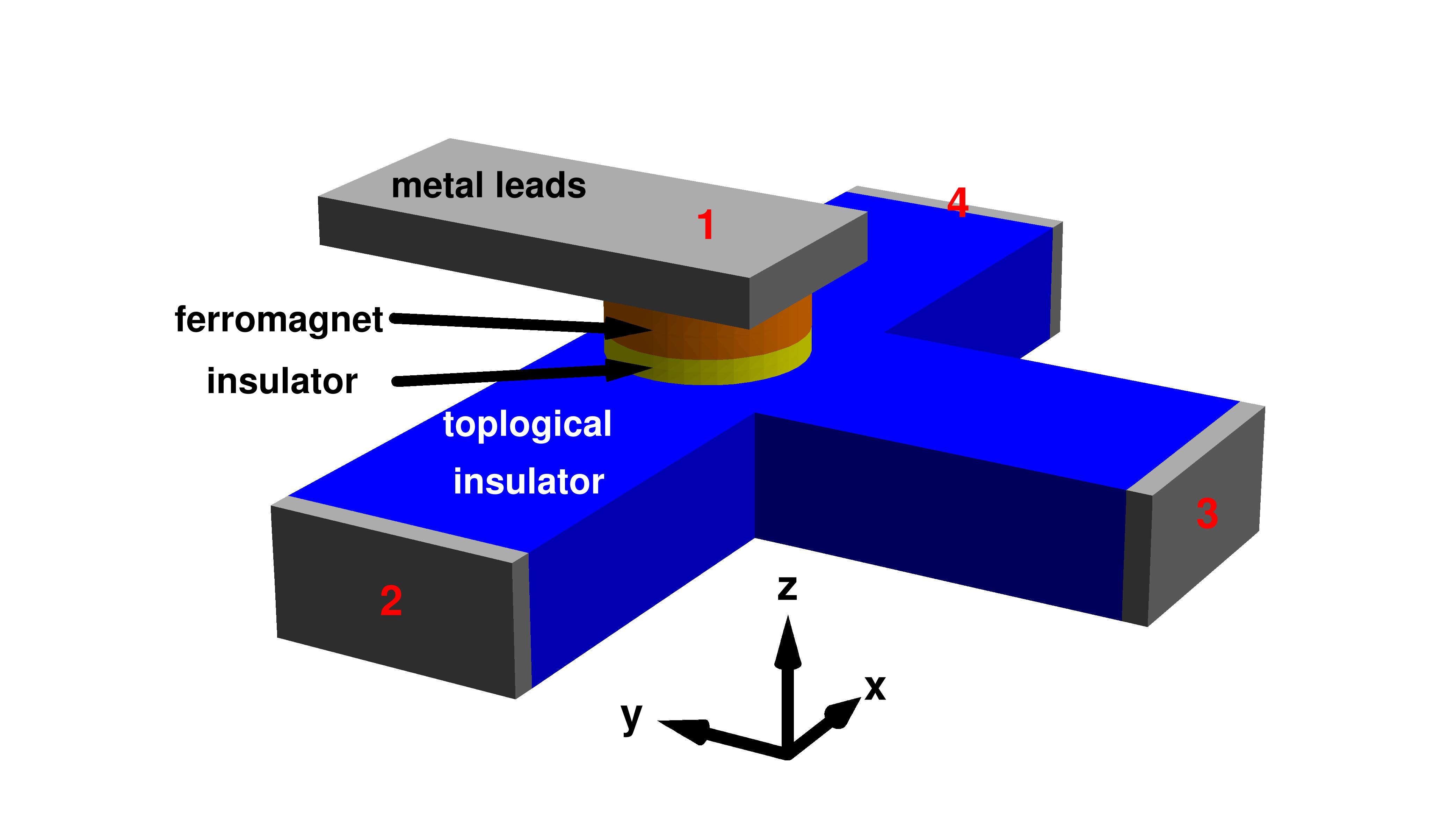}
\caption{Schematic of the spin Hall effect tunneling spectroscopy device used by Liu \textit{et al.}\cite{LLiu2}. An alternating current between lead 1 and 3 drives a spin polarized tunneling current into the surface states of the TI. An asymmetric charge accumulation due to the spin momentum locking of the surface states then leads to a voltage drop between lead 2 and 4.}
\label{schematic}
\end{figure}

A schematic of a Spin Hall effect tunneling spectroscopy device is shown in
figure~\ref{schematic}. The main idea is to create a spin polarized current by
a ferromagnet and feed it through leads 1 and 3. The spin texture
of the surface states then creates a transverse voltage between leads 2 and 4
which depends on the voltage applied between leads 1 and 3.
We will show in this work that the out-of-plane component of the spin
in the topological surface states can be obtained from a ratio of differential
conductances for out-of-plane polarization of the ferromagnet and in-plane
polarization of the ferromagnet.

\section{Results}
In this section we want to derive a method that allows to determine the
out-of-plane spin component of topological surface states \cite{GoetteDiss}. When a voltage is applied between lead 1 and 3 of the device shown in figure \ref{schematic}, a spin polarized current will tunnel from the ferromagnet into the surface states of the topological insulator. Due to the spin momentum locking of the surface states, the resulting currents towards the leads 2 and 4 are unequal, giving rise to an asymmetric charge accumulation and therefore a voltage $V_{\textrm{SH}}$ between these leads. The change of this voltage $\frac{dV_{\textrm{SH}}}{dI}$ is then proportional to the difference of the differential conductances (DC) $G\left(U\right)=\frac{dI\left(U\right)}{dU}$ with respect to the two leads. In the following we are only interested in relative DCs for different polarizations of the ferromagnet. The precise prefactor of the DC, which depends on microscopic details of the junction, drops out and thus plays no role here.
 
\subsection{Model}
We base our derivations on the realistic four band tight-binding Hamiltonian for the Bi$_{2}$Se$_{3}$ class of materials by Liu \textit{et al.}\cite{Liu-PRB2010}
\begin{equation}
H\left(\mathbf{k}\right)=\epsilon_{0}(\mathbf{k})\mathbb{I}_{4\times4}+\sum_{i=1}^{4}m_{i}\left(\mathbf{k}\right)\Gamma^{i}+\mathcal{R}_{1}\left(\mathbf{k}\right)\Gamma^{5}+\mathcal{R}_{2}\left(\mathbf{k}\right)\Gamma^{3}\label{eq:hamiltonian}
\end{equation}
with Dirac $\Gamma$ matrices
\begin{equation}
\Gamma^{1,2,3,4,5}=\left(\tau_{1}\otimes\sigma_{1},\tau_{1}\otimes\sigma_{2},\tau_{2}\otimes\mathbb{I}_{2\times2},\tau_{3}\otimes\mathbb{I}_{2\times2},\tau_{1}\otimes\sigma_{3}\right)\nonumber
\end{equation}
written in the basis $\left(1\uparrow,1\downarrow,2\uparrow,2\downarrow\right)$ and coefficients for a hexagonal lattice\cite{Hao-PRB2011}
\begin{eqnarray}
\epsilon_{0}(\mathbf{k}) & = & C_{0}+2C_{1}\left(1-\cos k_{z}\right)\nonumber \\
 &  & +\frac{4}{3}C_{2}\left(3-2\cos\frac{1}{2}k_{x}\cos\frac{\sqrt{3}}{2}k_{y}-\cos k_{x}\right)\nonumber \\
m_{1}(\mathbf{k}) & = & A_{0}\frac{2}{\sqrt{3}}\cos\frac{1}{2}k_{x}\sin\frac{\sqrt{3}}{2}k_{y}\nonumber \\
m_{2}(\mathbf{k}) & = & -A_{0}\frac{2}{3}\left(\sin\frac{1}{2}k_{x}\cos\frac{\sqrt{3}}{2}k_{y}+\sin k_{x}\right)\nonumber \\
m_{3}(\mathbf{k}) & = & B_{0}\sin k_{z}\nonumber \\
m_{4}(\mathbf{k}) & = & M_{0}+2M_{1}\left(1-\cos k_{z}\right)\nonumber \\
 &  & +\frac{4}{3}M_{2}\left(3-2\cos\frac{1}{2}k_{x}\cos\frac{\sqrt{3}}{2}k_{y}-\cos k_{x}\right)\nonumber
\end{eqnarray}

The hexagonal warping along with the out-of-plane tilt of the surface state
spin is given by the coefficients $\mathcal{R}_{1}$ and $\mathcal{R}_{2}$ 
which are at least of third order in momentum $\mathbf{k}$\cite{Goette-PRA2014}
\begin{eqnarray}
\mathcal{R}_{1}\left(\mathbf{k}\right) & = & 2R_{1}\left(\cos\sqrt{3}k_{y}-\cos k_{x}\right)\sin k_{x}\nonumber \\
\mathcal{R}_{2}\left(\mathbf{k}\right) & = & \frac{16}{3\sqrt{3}}R_{2}\left(\cos\frac{\sqrt{3}}{2}k_{y}-\cos\frac{3}{2}k_{x}\right)\sin\frac{\sqrt{3}}{2}k_{y}.\nonumber
\end{eqnarray}
We will test our method numerically for two different sets of parameters valid for two different topological insulators.\cite{Liu-PRB2010} For Bi$_2$Se$_3$ we have the lattice constants $a=4.14\textrm{\AA}$
and $c=\frac{28.64}{15}\textrm{\AA}$\cite{Madelung1998} and parameters $A_{0}=0.804\textrm{eV}$,
$B_{0}=1.184\textrm{eV}$, $C_{1}=1.575\textrm{eV}$, $C_{2}=1.774\textrm{eV}$,
$M_{0}=-0.28\textrm{eV}$, $M_{1}=1.882\textrm{eV}$, $M_{2}=2.596\textrm{eV}$,
$R_{1}=0.713\textrm{eV}$, and $R_{2}=-1.597\textrm{eV}$. For Sb$_2$Te$_3$ we have the lattice constants $a=4.25\textrm{\AA}$
and $c=\frac{30.35}{15}\textrm{\AA}$\cite{Madelung1998} and parameters $A_{0}=0.8\textrm{eV}$,
$B_{0}=0.415\textrm{eV}$, $C_{1}=-3.027\textrm{eV}$, $C_{2}=-0.597\textrm{eV}$,
$M_{0}=-0.22\textrm{eV}$, $M_{1}=4.797\textrm{eV}$, $M_{2}=2.986\textrm{eV}$,
$R_{1}=1.344\textrm{eV}$, and $R_{2}=-3.187\textrm{eV}$. The small energy shift $C_0$ is set to zero in both cases .
\subsubsection{Derivation}
In this subsection we derive an approximate formula, which allows to calculate
the out-of-plane spin component from experimental conductance data (equation~\ref{eq:q_formula}). In the
next subsection we will then test this formula and show that the extracted
values for the out-of-plane polarization compare very well with the actual values of the two
sets of parameters for Bi$_2$Se$_3$ and Sb$_2$Te$_3$.

When we expand equation \ref{eq:hamiltonian} up to second order in momentum $\mathbf{k}$, we can derive an analytical expression for the surface states in the top layer of the TI\cite{Goette-PRA2014}
\begin{equation}
\psi_{\pm}\left(\varphi\right)=\frac{1}{\sqrt{2}}\left(\begin{array}{c}
{\pm}u_{-}e^{-i\left(\varphi-\frac{\pi}{2}\right)}\\
u_{-}\\
{\pm}u_{+}e^{-i\left(\varphi-\frac{\pi}{2}\right)}\\
u_{+}
\end{array}\right),
\end{equation}
where $\pm$ is for the upper and lower Dirac cone respectively, $u_{\pm}=\sqrt{\frac{M_{1}{\pm}C_{1}}{2M_{1}}}$ and $\varphi$ is the in-plane polar angle of the momentum. The eigenenergies of these states are isotropic, i.e. they depend only on $k=\sqrt{k_x^2+k_y^2}$ but not on $\varphi$
\begin{equation}
E_{\pm}=-\frac{C_{1}M_{0}}{M_{1}}+\left(C_{2}-\frac{C_{1}}{M_{1}}M_{2}\right)k^{2}\pm A_{0}\sqrt{1-\frac{C_{1}^{2}}{M_{1}^{2}}}k.
\end{equation}
The spin of these states lies within the surface plane and is always perpendicular to the in-plane momentum. In order to keep things analytically solvable, we make a simplified assumption when we add the out-of-plane tilt of the spin\begin{equation}
\bar{\psi}_{\pm}\left(\varphi\right)=\frac{1}{\sqrt{2}}\left(\begin{array}{c}
{\pm}u_{-}\sqrt{1{\pm}q_{0}\cos3\varphi}e^{-i\left(\varphi-\frac{\pi}{2}\right)}\\
u_{-}\sqrt{1{\mp}q_{0}\cos3\varphi}\\
{\pm}u_{+}\sqrt{1{\pm}q_{0}\cos3\varphi}e^{-i\left(\varphi-\frac{\pi}{2}\right)}\\
u_{+}\sqrt{1{\mp}q_{0}\cos3\varphi}
\end{array}\right).\label{eq:states_q0}
\end{equation}
Here, $q_{0}$ is the energy dependent out-of-plane spin polarization that is
modulated by $\cos3\varphi$ to account for the hexagonal warping. The in-plane
spin component remains perpendicular to the in-plane momentum. We further
simplify equation \ref{eq:states_q0} by making the replacement
$q_{0}\cos3\varphi\rightarrow\bar{q}\textrm{sign}\left(\cos3\varphi\right)$,
with the mean value
$\bar{q}=\frac{3q_{0}}{\pi}\int^{\frac{\pi}{6}}_{-\frac{\pi}{6}}d\varphi\cos3\varphi$. This
approximation allows us to obtain analytical expressions for the differential
conductance below.
\\
The states of the ferromagnetic electrode are obtained by reducing
equation \ref{eq:hamiltonian} to a single orbital and using $C_1=C_2=0.25eV$ and $C_0=-0.75eV$. All other parameters are set to 0. The spatial dependence perpendicular to the surface is then given by the superposition of an incoming and outgoing plane wave $\sin\left(zk_z\right)$. We can write down states for the surface layer ($z=1$) with a definite polarization either within the surface plane defined by the polar angle $\varphi_F$
\begin{equation}
\psi_{\varphi_F}\left(k_z\right)=\frac{1}{\sqrt{2}}\sin{k_z}\left(\begin{array}{c}
e^{-i\varphi_F}\\
1
\end{array}\right)\label{eq:states_FMxy}
\end{equation}
or out-of-plane
\begin{equation}
\psi_{z\uparrow}\left(k_z\right)=\sin{k_z}\left(\begin{array}{c}
1\\
0
\end{array}\right), 
\psi_{z\downarrow}\left(k_z\right)=\sin{k_z}\left(\begin{array}{c}
0\\
1
\end{array}\right).
\label{eq:states_FMz}
\end{equation}
The tunneling Hamiltonian of the insulating barrier is given by
\begin{equation}
H_{T}=- C_B \sum_{k_x,k_y,\alpha.\sigma} d^\dagger_{k_x,k_y,\alpha.\sigma}
c_{k_x,k_y,\sigma} + \mathrm{h.c.} \, ,\label{eq:thamilton}
\end{equation}
where $d^\dagger_{k_x,k_y,\alpha.\sigma}$ creates an electron in the top layer of the topological insulator and $c_{k_x,k_y,\sigma}$
destroys an electron in the bottom layer of the
ferromagnet. $\alpha$ and $\sigma$ are orbital and spin quantum numbers and
$C_B$ is the hopping matrix element.
Using Fermi's golden Rule
\begin{equation}
\Gamma_{mn}=\frac{2\pi}{\hbar}\delta\left(E_{n}-E_{m}\right)\left|\left\langle n\left|H_{T}\right|m\right\rangle \right|^{2},
\end{equation}
for the transition rate from an initial state $m$ into a final state $n$ we get transition probabilities
\begin{eqnarray}
\left|\left\langle \psi_{\varphi_{F}}\left|H_{T}\right|\bar{\psi}_{\pm}\right\rangle \right|^{2} & = & \frac{1}{2}C_{B}^{2}\left(u_{-}^{2}+u_{+}^{2}\right)\sin^{2}k_{z}(k)\nonumber\\
&&\cdot\left[1\mp\sqrt{1-\bar{q}^{2}}\sin\left(\varphi_{F}-\varphi\right)\right]\,,\label{eq:mat_elem_FMxy}
\end{eqnarray}
\begin{eqnarray}
\left|\left\langle \psi_{z\uparrow}\left|H_{T}\right|\bar{\psi}_{\pm}\right\rangle \right|^{2} & = & \frac{1}{2}C_{B}^{2}\left(u_{-}^{2}+u_{+}^{2}\right)\sin^{2}k_{z}(k)\\
&& \cdot\left(1\pm\bar{q}\text{sign}\left(\cos3\varphi\right)\right)\,,\label{eq:mat_zup}
\end{eqnarray}
and
\begin{eqnarray}
\left|\left\langle \psi_{z\downarrow}\left|H_{T}\right|\bar{\psi}_{\pm}\right\rangle \right|^{2} & = & \frac{1}{2}C_{B}^{2}\left(u_{-}^{2}+u_{+}^{2}\right)\sin^{2}k_{z}(k)\\
&& \cdot\left(1\mp\bar{q}\text{sign}\left(\cos3\varphi\right)\right)\,.\label{eq:mat_zdown}
\end{eqnarray}
These probabilities can be inserted into the DC\cite{Goette-PRA2014}
\begin{eqnarray}
G_F\left(T,U\right) & = & \frac{\textrm{const.}}{T}\int_{0}^{k_{0}}dkk\int_{-\pi}^{\pi}d\varphi\Bigg(f\left(\varphi\right)\frac{\left|\left\langle \psi_{F}\left|H_{T}\right|\bar{\psi}_{+}\right\rangle \right|^{2}}{\cosh^{2}\left(\frac{E_{+}-eU}{2k_{B}T}\right)}\nonumber \\
 &  & +f\left(\varphi-\pi\right)\frac{\left|\left\langle \psi_{F}\left|H_{T}\right|\bar{\psi}_{-}\right\rangle \right|^{2}}{\cosh^{2}\left(\frac{E_{-}-eU}{2k_{B}T}\right)}\Bigg)\,.\label{eq:G-_mat_el-1}
\end{eqnarray}
where $U$ is the bias voltage over the barrier, $T$ is the temperature, $\psi_{F}$ is an arbitrary ferromagnetic state, and
$f\left(\varphi\right)$ gives the probability that an electron that starts its
propagation in the surface of the TI at an angle $\varphi$ ends up at a certain electrode. When we introduce new abbreviations $A_{\pm}\left(k\right)=\frac{C_{B}^{2}\sin^{2}k_{z}(k)}{\cosh^{2}\left(\frac{E_{\pm}-eU}{2k_{B}T}\right)}\left(u_{-}^{2}+u_{+}^{2}\right)$,
equal for all $\psi_{F}$, and $g_{\pm}^{F}\left(\varphi\right)$ containing
the remaining parts, we can rewrite equation \ref{eq:G-_mat_el-1} as follows
\begin{eqnarray}
G_{F}\left(T,U\right) & = & \frac{\textrm{const.}}{T}\int_{0}^{k_{0}}dkk\int_{-\pi}^{\pi}d\varphi(f\left(\varphi\right)g_{+}^{F}\left(\varphi\right)A_{+}\left(k\right)\nonumber \\
& & +f\left(\varphi-\pi\right)g_{-}^{F}\left(\varphi\right)A_{-}\left(k\right))\,.\label{eq:G_gA_1}
\end{eqnarray}
$g_{\pm}^{F}\left(\varphi\right)$ depends on the ferromagnetic states, but always satisfies
$g_{-}^{F}\left(\varphi+\pi\right)=g_{+}^{F}\left(\varphi\right)$. 
As $\bar{q}\left(E\right)$ varies slowly as a function of energy and the temperature is low, we can replace $\bar{q}\left(E\right)$ with $\bar{q}\left(U\right)$ and take it out of the integral. When we further assume that $f\left(\varphi\right)=f\left(\varphi+2\pi\right)=f\left(-\varphi\right)$, we can simplify equation \ref{eq:G_gA_1}
\begin{eqnarray}
G_{F}\left(T,U\right) & = & \frac{\textrm{const.}}{T}\int_{0}^{k_{0}}dkk\int_{-\pi}^{\pi}d\varphi f\left(\varphi\right)(g_{+}^{F}\left(\varphi\right)A_{+}\left(k\right)\nonumber \\
 & & +g_{-}^{F}\left(\varphi+\pi\right)A_{-}\left(k\right))\nonumber \\
 & = & \frac{\textrm{const.}}{T}\int_{0}^{k_{0}}dkk\left(A_{+}\left(k\right)+A_{-}\left(k\right)\right)\nonumber \\
 & & \int_{-\pi}^{\pi}d\varphi f\left(\varphi\right)g_{+}^{F}\left(\varphi\right)\nonumber \\
 & = & G_{0}\left(T,U\right)\int_{-\pi}^{\pi}d\varphi f\left(\varphi\right)g_{+}^{F}\left(\varphi\right)\,.
\end{eqnarray}
Here,
$G_{0}\left(T,U\right)=\frac{\textrm{const.}}{T}\int_{0}^{k_{0}}dkk\left(A_{+}\left(k\right)+A_{-}\left(k\right)\right)$
is a term, which is independent of the ferromagnetic state $\psi_F$.
The DC with respect to the opposite electrode is obtained by the substitution $f\left(\varphi\right)\rightarrow f\left(\varphi+\pi\right)$. So the difference of the DCs with respect to the two electrodes is
\begin{equation}
\Delta G_{F}\left(T,U\right)=G_{0}\left(T,U\right)\int_{-\pi}^{\pi}d\varphi f\left(\varphi\right)\left(g_{+}^{F}\left(\varphi\right)-g_{-}^{F}\left(\varphi\right)\right)\,.
\end{equation}
Considering now the device in figure \ref{schematic}, all electrons that initially move in positive x-direction end up at the corresponding electrode, so
\begin{equation}
f\left(\varphi\right)=\begin{cases}
1 & \textrm{for}\,\varphi \in \left[-\frac{\pi}{2},\frac{\pi}{2}\right]\\
0 & \textrm{else}.
\end{cases}\label{eq:f3D-1}
\end{equation}
With this we obtain the following $\Delta G$
for a ferromagnet fully polarized in-plane
\begin{equation}
\Delta G_{\varphi_{F}}\left(T,U,\varphi_{F}\right)=G_{0}\left(T,U\right)2\sqrt{1-\bar{q}^{2}\left(U\right)}\sin\varphi_{F}\,,
\end{equation}
where the polar angle $\varphi_{F}$ has to be adjusted to the in-plane
polarization angle of the TI surface states that propagate along the x-axis,
i.e. $\varphi=\frac{\pi}{2}$. For a ferromagnet fully polarized out-of-plane
we find
\begin{equation}
\Delta G_{z\uparrow,\downarrow}\left(T,U\right)=\pm G_{0}\left(T,U\right)\frac{\pi}{3}\bar{q}\left(U\right)\,.
\end{equation}
For a ferromagnet with finite polarization, we simply have to take a weighted sum of terms with opposite polarization. So, we finally get
\begin{equation}
\Delta G_{\text{ip}}\left(T,U,\varphi\right)=G_{0}\left(T,U\right)2\sqrt{1-\bar{q}^{2}\left(U\right)}\sin\varphi_{F}\Delta n_{\text{ip}}\label{eq:dG_ip}
\end{equation}
for the in-plane polarization and
\begin{equation}
\Delta G_{\text{op}}\left(T,U,\varphi\right)=G_{0}\left(T,U\right)\frac{\pi}{3}\bar{q}\left(U\right)\Delta n_{\text{op}}\label{eq:dG_op}
\end{equation}
for the out-of-plane polarization, where $\Delta n=n_{+}-n_{-}$, with $n_{+}+n_{-}=1$, is the relative density of states of spin-up and spin-down states. When we take the ratio of equation \ref{eq:dG_ip} and equation \ref{eq:dG_op}, $G_0$ cancels out and we can solve for the mean out-of-plane polarization of the TI surface states
\begin{equation}
\bar{q}\left(U\right)=\sqrt{\frac{1}{1+\left(\frac{\Delta G_{\text{ip}}}{\Delta G_{\text{op}}}\frac{\Delta n_{\text{op}}}{\Delta n_{\text{ip}}}\frac{\pi}{6\sin\varphi_{F}}\right)^{2}}}.\label{eq:q_formula}
\end{equation}
This equation allows to obtain the out-of-plane spin polarization from the
ratio of the differential conductances for in-plane and out-of-plane
polarization of the ferromagnet.

\begin{figure*}[t]%
\subfloat[Bi$_2$Se$_3$]{%
\includegraphics*[width=.5\textwidth]{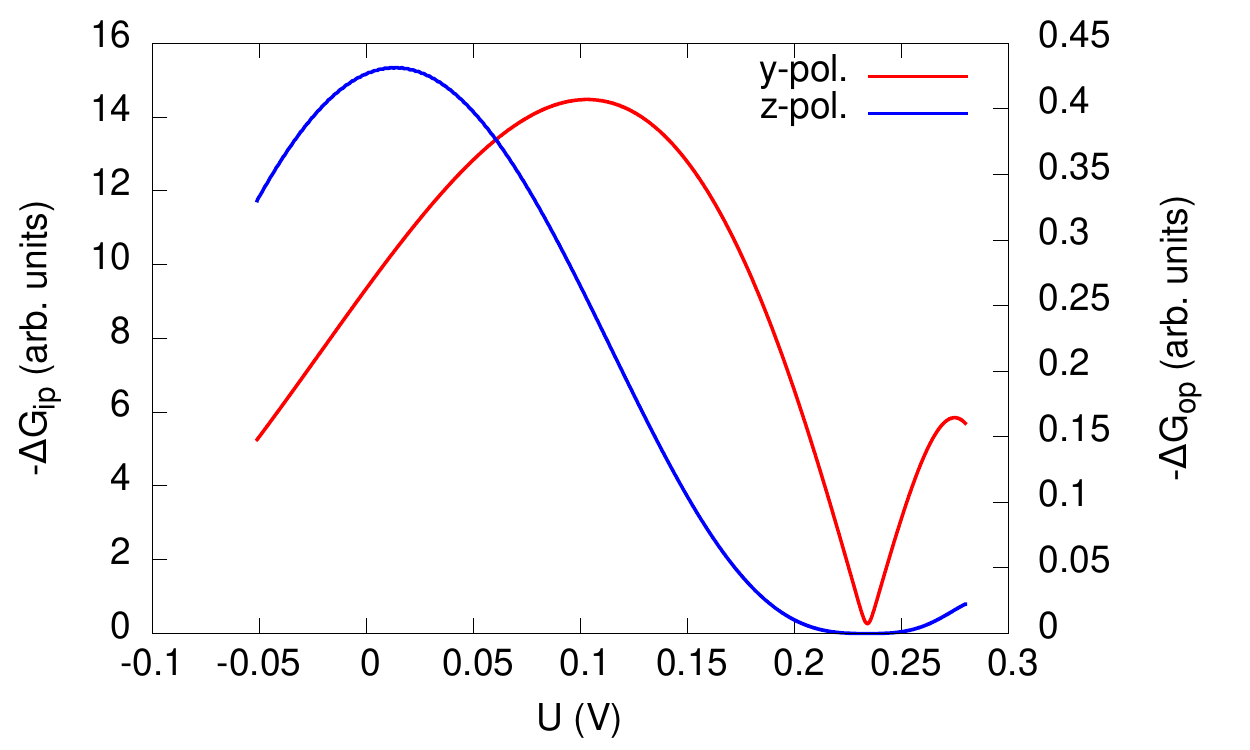}}\hfill
\subfloat[Sb$_2$Te$_3$]{%
\includegraphics*[width=.5\textwidth]{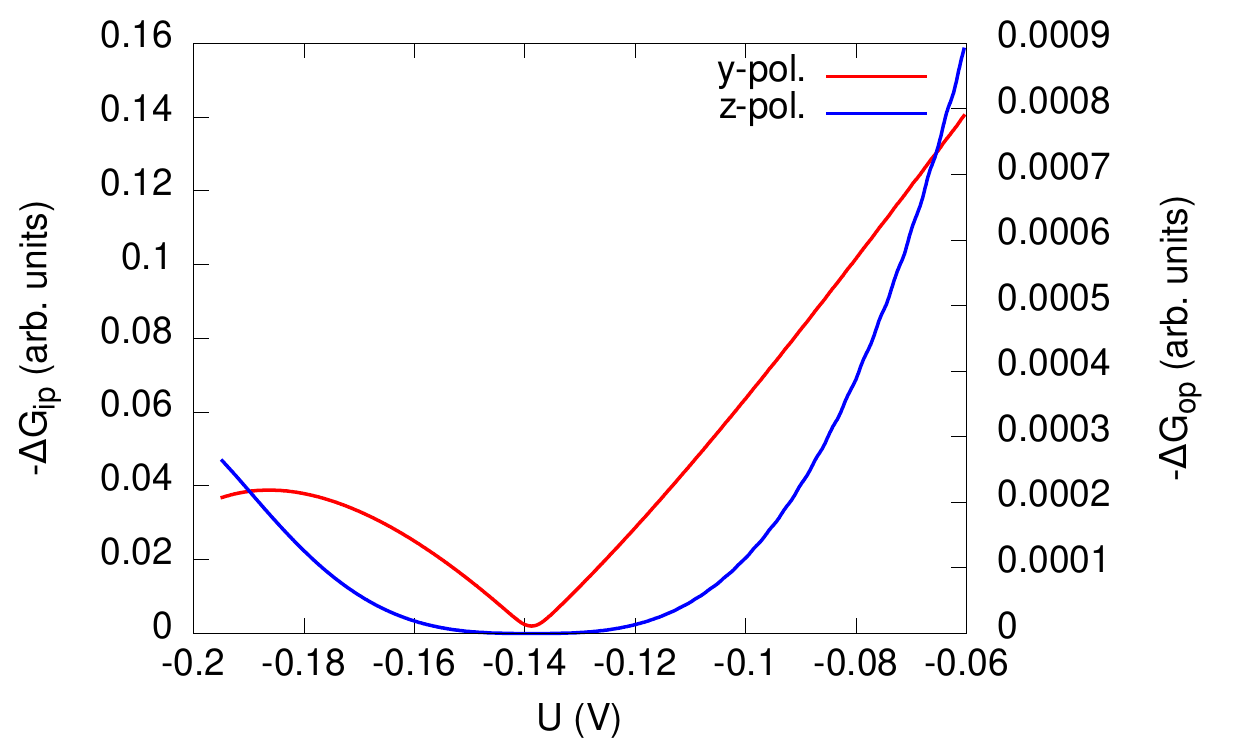}}\hfill
\caption{%
Differential conductance differences $\Delta G$ as a function of bias voltage U, calculated from numerical eigenstates of the full Hamiltonian, at 10K for energy values inside the bulk gap. For both materials, $\Delta G_{\textrm{ip}}$ (y-polarization) shows a sharp minimum at the Dirac point. The minimum of $\Delta G_{\textrm{op}}$ (z-polarization) is much shallower. Note the different scales for in-plane and out-of-plane curves.}
\label{DC}
\end{figure*}

Under the assumption that the approximation of the out-of-plane polarization
\begin{equation}
q\left(U,\varphi\right)=\pm q_{0}\left(U\right)\cos3\varphi\label{eq:q_approx}
\end{equation}
of the TI surface states is good, the angular dependence can be calculated from $\bar{q}\left(U\right)$
\begin{eqnarray}
q\left(U,\varphi\right) & = & \pm\bar{q}\left(U\right)\frac{\pi}{3}\frac{\cos3\varphi}{\int_{-\pi/6}^{\pi/6}d\varphi^{\prime}\cos3\varphi^{\prime}}\nonumber\\
 & = &\pm\bar{q}\left(U\right)\frac{\pi}{2}\cos3\varphi\,.\label{eq:q_of_E_phi}
\end{eqnarray}
However, as only the alternating sign of the out-of-plane polarization was considered in the above derivation, any other valid angular dependence could be used here as well. 

\subsection{Numerical test}
\begin{figure}[t]%
\includegraphics*[width=\linewidth]{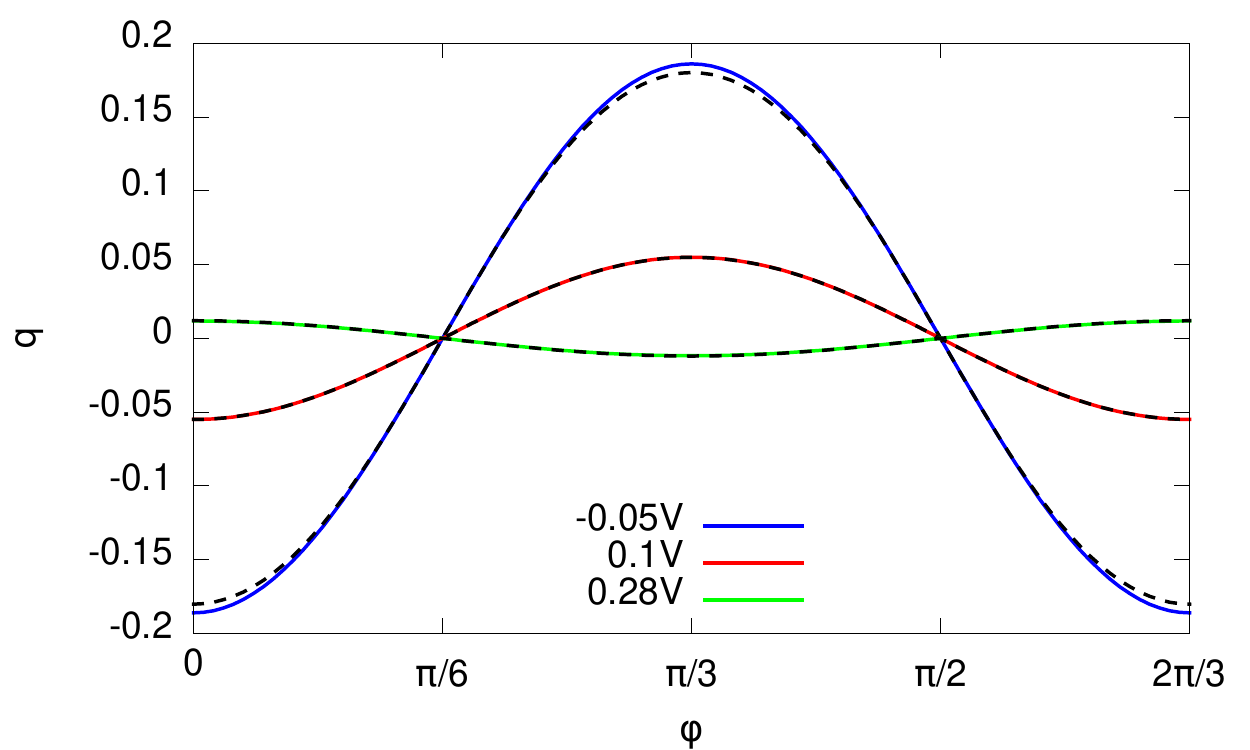}
\caption{Dependence of the out-of-plane polarization $q\left(U\right)$ of Bi$_2$Se$_3$ on the in-plane polar angle $\varphi$ of the momentum for three representative bias voltages. The solid colored lines are derived from the numerical mean polarization values and equation \ref{eq:q_of_E_phi} and the dashed black lines show equation \ref{z-expectation} evaluated along a constant energy contour. Apart from a small amplitude deviation for $U=-0.05\textrm{V}$ the agreement is very good.}
\label{phidependence}
\end{figure}

In this section we want to test our result, equation \ref{eq:q_formula}, as well as
our approximation, equation \ref{eq:q_of_E_phi}, by comparison to results based on
numerical eigenvalues of the full Hamiltonian equation \ref{eq:hamiltonian}. The
eigenstates of the ferromagnet, which for simplicity is assumed to be fully polarized, are still described by the analytical formulas equation \ref{eq:states_FMxy} and \ref{eq:states_FMz}. As the numerical eigenstates are only given for discrete momenta, we have to replace the integrals in equation \ref{eq:G-_mat_el-1} by a sum over all numerical TI eigenstates and analytical ferromagnet states that fulfill energy and in-plane momentum conservation.\cite{Goette-PRA2014} For the numerical states we use a hexagonal in-plane momentum discretization of $\frac{2}{\sqrt{3}}\frac{2\pi}{2000}$ and 50 (Bi$_2$Se$_3$) respectively 200 (Sb$_2$Te$_3$) real space lattice sites in z-direction. Propagation directions are obtained from a difference quotient for small variations of the in-plane momenta. The resulting $\Delta G$ at 10K within the TI bulk gaps are shown in figure \ref{DC}.\\
For the in-plane polarization, both materials show a sharp minimum at the
Dirac point with an approximately linear increase towards higher and lower
energies. The minimum for the out-of-plane polarization is shallower because
the out-of-plane spin component of the surface states decreases to zero
towards the Dirac point. Away from the Dirac point $\Delta G_{\textrm{op}}$ is
much steeper because out-of-plane spin and density of states both increase at the same time. Towards the bulk gap edges, most of the curves decrease. This is consistent with experiments \cite{Alpichshev2010} and can be mostly attributed to the increased decay length of the surface states as they merge into bulk states, leading to decreased transistion matrix elements. As the surface states of Sb$_2$Te$_3$ merge with the conduction bands at a much higher energy and only the energy range of the bulk gap is shown here, the decrease is not visible there. The $\Delta G_{\textrm{ip}}$ curves are further decreased by the increasing out-of-plane tilt of the surface state polarization. Again, the effect is smaller for Sb$_2$Te$_3$ because the hexagonal deformation of the Dirac cone and out-of-plane tilt of the surface state spin is smaller. The overall smaller values of $\Delta G_{\textrm{op}}$ compared to $\Delta G_{\textrm{ip}}$ come from the generally smaller out-of-plane spin component and the fact that two-thirds of the surface states compensate each other for an out-of-plane tunneling current due to the alternating sign of the out-of-plane spin.\\
\begin{figure*}[t]%
\subfloat[Bi$_2$Se$_3$]{%
\includegraphics*[width=.5\textwidth]{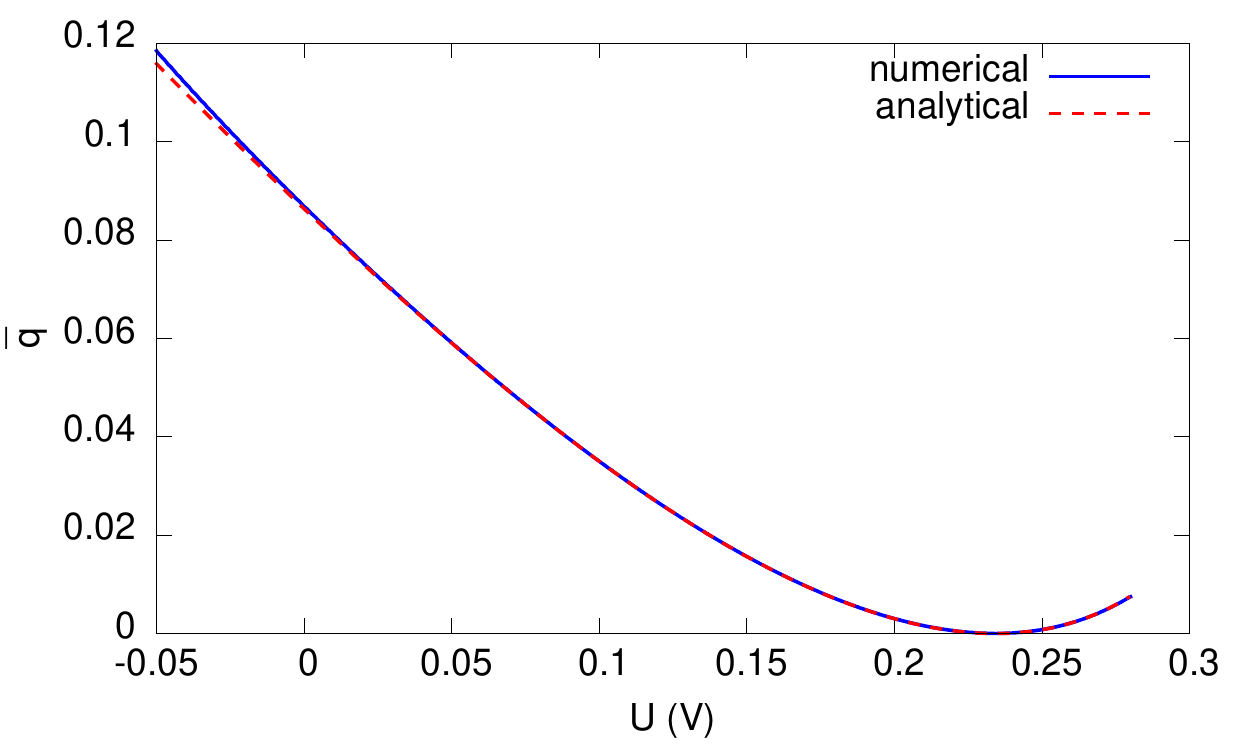}}\hfill
\subfloat[Sb$_2$Te$_3$]{%
\includegraphics*[width=.5\textwidth]{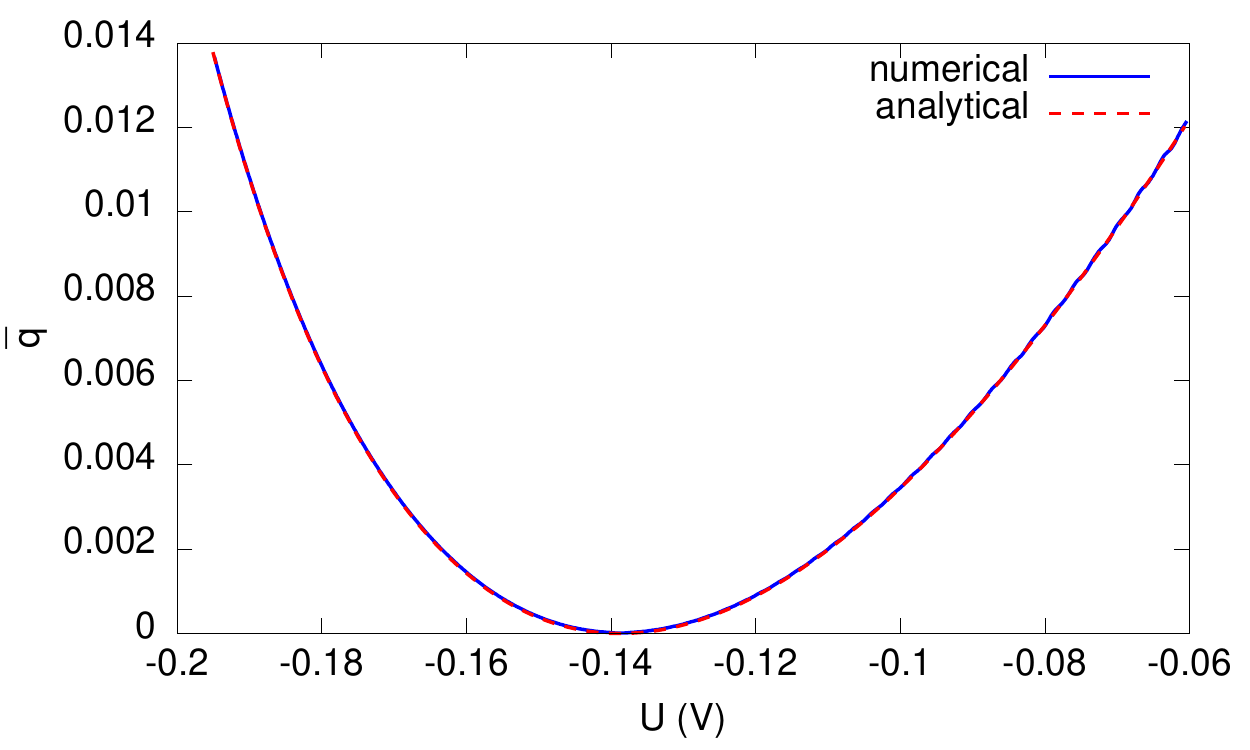}}\hfill
\caption{%
Comparison of the mean out-of-plane spin polarization $\bar{q}$ (equation \ref{eq:q_formula}) derived from numerical $\Delta G$ curves in figure \ref{DC} and the mean value of the analytical expectation value equation \ref{z-expectation}. There is only a small deviation of the two curves at the lower bulk gap edge of Bi$_2$Se$_3$.}
\label{meanpolarization}
\end{figure*}
Next we calculate the mean out-of-plane spin from the curves in figure
\ref{DC} with parameters $\Delta n_{\textrm{ip}}=\Delta n_{\textrm{op}}=1$ and
$\varphi_{F}=\frac{\pi}{2}$. The results are presented in figure
\ref{meanpolarization} (solid lines). For comparison we show the mean value
$\bar{q}_{\textrm{ana}}\left(E\right)$ of the absolute value of the z-spin
expectation value
\begin{equation}
p_{z}=\psi_{\pm}^{\prime\dagger}\left(\Sigma_{z}\psi_{\pm}^{\prime}\right)=\pm\frac{R_1}{\sqrt{R_1^2+m_1^2+m_2^2}}\label{z-expectation}
\end{equation}
of the analytical surface states $\psi_{\pm}^{\prime}$ involving the full
in-plane momentum dependence (dashed lines).\cite{Goette-PRA2014} Except for a small deviation for Bi$_2$Se$_3$ at the lower bulk gap edge, the numerical and analytical curves agree very well. The small deviation is likely due to the replacement $q_{0}\cos3\varphi\rightarrow\bar{q}\textrm{sign}\left(\cos3\varphi\right)$ which leads to a changed angular dependent weighting of the matrix elements for tunneling with in-plane spin which causes an overestimate of $\Delta G_{\textrm{ip}}$. Besides, the hexagonal warping of the Fermi surface, which was neglected in the derivation of equation \ref{eq:q_formula}, might have a small effect.\\
Figure \ref{phidependence} shows a comparison of the real analytical angular
dependence of the out-of-plane spin and the approximation equation \ref{eq:q_of_E_phi} for three representative energy values for Bi$_2$Se$_3$. The agreement is again very good and therefore validates the assumption for the wavefunction in equation \ref{eq:states_q0}. There is only a small amplitude deviation for the lowest energy value due to the overestimate of $\bar{q}$. 

\section{Summary}
We derived an analytical formula that allows to obtain the out-of-plane spin
polarization of topological surface states from spin Hall effect tunneling
spectra.
We have tested this formula by applying it to full numerical calculations 
of the differential conductances using two different sets of parameters
appropriate for Bi$_2$Se$_3$ and Sb$_2$Te$_3$. The extracted out-of-plane spin
polarization was shown to be in very good agreement with the actual out-of-plane spin
polarization of the two sets of parameters. This shows that a reliable
extraction of the out-of-plane spin polarization by spin Hall effect tunneling
spectroscopy is feasible. Together with the Meservey-Tedrow technique, which
allows a reliable measurement of the in-plane spin polarization, these two
techniques can provide a detailed measurement of the spin texture of
topological surface states based on tunneling spectroscopy.

\begin{acknowledgement}
Financial support from the German Research Foundation (DFG) via priority program SPP~1666 ``Topological Insulators'' is gratefully acknowledged. We would like to thank G.~Reiss for valuable discussions.

\end{acknowledgement}

%
%

\providecommand{\WileyBibTextsc}{}
\let\textsc\WileyBibTextsc
\providecommand{\othercit}{}
\providecommand{\jr}[1]{#1}
\providecommand{\etal}{~et~al.}

\end{document}